\documentclass{elsart}
\begin{document}
\begin{frontmatter}
\title{
The Role of Ground State Correlations in the Single-Particle Strength of
Odd Nuclei with Pairing.
}
\author[Obninsk]{A.V. Avdeenkov\thanksref{RFFR}}
\author[Obninsk]{S.P. Kamerdzhiev\thanksref{RFFR}}
\address[Obninsk]
{State Scientific Centre Institute of Physics and  Power Engineering,\\
249020 Obninsk, Kaluga region, sq. Bondarenko 1, Russia}
\thanks[RFFR]{
Partially supported by the Russian Funds for Fundamental
Research, grant No.96-02-17250.
}

\begin{abstract}
A method based on the consistent use of the Green
function formalism has been developed to calculate the distribution of the
single-particle strength in odd nuclei with pairing. The method takes into
account the quasiparticle-phonon interaction, ground state correlations
and a "refinement" of phenomenological single-particle energies and pairing
gap values from the quasiparticle-phonon interaction under consideration.
The calculations for $^{121}Sn$ and $^{119}Sn$ that were performed in
the quasiparticle$\otimes$phonon approximation, have shown a reasonable
agreement with experiment. The ground state correlations play a noticeable role
and mostly improve the agreement with experiment or shift the results to the
right direction.
\end{abstract}
\begin{keyword}
 Single-particle strength, ground state correlations, pairing,
Green functions.
\end{keyword}
\end{frontmatter}

\section{Introduction}
As is well known the role of the quasiparticle-phonon interaction is
essential for the description of excitations in odd-mass nuclei ~\cite{r1,r2,r3}.
In magic~\cite{r1} and semi- magic~\cite{r4} nuclei it is possible to restrict ourselves to
the approximation of the squared phonon creation amplitude $g^{2}$ in the
propagators of the integral equations under study ~\cite{r4}. In other
words, it is necessary to take into account at least the complex configurations
1p$\otimes$phonon or 1h$\otimes$phonon (for non-magic nuclei we will use the
unified notation 1qp$\otimes$phonon). In the Green function (GF) language
this means that for magic nuclei it is necessary to solve the Dyson equation
with the mass operator
\begin{eqnarray}
\label{M}
\unitlength=1.00mm
\special{em:linewidth 0.4pt}
\linethickness{0.4pt}
\begin{picture}(50.00,15.00)(-30,10)
\put(24.00,10.00){\line(1,0){20.00}}
\put(28.00,11.00){\circle{2.00}}
\put(40.00,11.00){\circle{2.00}}
\put(29.00,13.00){\circle*{1.00}}
\put(30.00,14.00){\circle*{1.0}}
\put(31.00,15.00){\circle*{1.0}}
\put(32.00,15.50){\circle*{1.00}}
\put(33.00,16.00){\circle*{1.00}}
\put(34.00,16.20){\circle*{1.00}}
\put(35.00,16.00){\circle*{1.00}}
\put(36.00,15.50){\circle*{1.00}}
\put(37.00,15.00){\circle*{1.00}}
\put(38.00,14.00){\circle*{1.00}}
\put(39.00,13.00){\circle*{1.00}}
\put(20.00,10.00){\line(1,0){5.00}}
\put(44.00,10.00){\line(1,0){5.00}}
\put(22.00,12.00){\line(2,-1){4.00}}
\put(26.00,10.00){\line(-2,-1){4.00}}
\put(32.00,12.00){\line(2,-1){4.00}}
\put(36.00,10.00){\line(-2,-1){4.00}}
\put(42.00,12.00){\line(2,-1){4.00}}
\put(46.00,10.00){\line(-2,-1){4.00}}
\put(10.00,10.00){\makebox(0,0)[cc]{\large M $=$}}
\end{picture}
\end{eqnarray}
where the circle is the phonon creation amplitude g.

The main difference of non-magic nuclei from magic ones is the necessity to take
into account Cooper pairing in the nuclear ground state. This means that we
should also consider the following simplest energy-dependent "anomalous" mass operators
~\cite{r4,r5}:
\begin{eqnarray}
\label{M1}
\unitlength=1.00mm
\special{em:linewidth 0.4pt}
\linethickness{0.4pt}
\begin{picture}(85.00,15.00)(-20,10)
\put(4.00,10.00){\line(1,0){20.00}}
\put(8.00,11.00){\circle{2.00}}
\put(20.00,11.00){\circle{2.00}}
\put(9.00,13.00){\circle*{1.00}}
\put(10.00,14.00){\circle*{1.0}}
\put(11.00,15.00){\circle*{1.0}}
\put(12.00,15.50){\circle*{1.00}}
\put(13.00,16.00){\circle*{1.00}}
\put(14.00,16.20){\circle*{1.00}}
\put(15.00,16.00){\circle*{1.00}}
\put(16.00,15.50){\circle*{1.00}}
\put(17.00,15.00){\circle*{1.00}}
\put(18.00,14.00){\circle*{1.00}}
\put(19.00,13.00){\circle*{1.00}}
\put(00.00,10.00){\line(1,0){5.00}}
\put(24.00,10.00){\line(1,0){5.00}}
\put(2.00,12.00){\line(2,-1){4.00}}
\put(6.00,10.00){\line(-2,-1){4.00}}
\put(9.00,12.00){\line(2,-1){4.00}}
\put(13.00,10.00){\line(-2,-1){4.00}}
\put(15.00,10.00){\line(2,-1){4.00}}
\put(15.00,10.00){\line(2,1){4.00}}
\put(24.00,10.00){\line(2,1){4.00}}
\put(24.00,10.00){\line(2,-1){4.00}}
\put(-10.00,10.00){\makebox(0,0)[cc]{\large $M^{(1)} =$}}
\put(64.00,10.00){\line(1,0){20.00}}
\put(68.00,11.00){\circle{2.00}}
\put(80.00,11.00){\circle{2.00}}
\put(69.00,13.00){\circle*{1.00}}
\put(70.00,14.00){\circle*{1.0}}
\put(71.00,15.00){\circle*{1.0}}
\put(72.00,15.50){\circle*{1.00}}
\put(73.00,16.00){\circle*{1.00}}
\put(74.00,16.20){\circle*{1.00}}
\put(75.00,16.00){\circle*{1.00}}
\put(76.00,15.50){\circle*{1.00}}
\put(77.00,15.00){\circle*{1.00}}
\put(78.00,14.00){\circle*{1.00}}
\put(79.00,13.00){\circle*{1.00}}
\put(60.00,10.00){\line(1,0){5.00}}
\put(84.00,10.00){\line(1,0){5.00}}
\put(63.00,10.00){\line(2,1){4.00}}
\put(63.00,10.00){\line(2,-1){4.00}}
\put(75.00,12.00){\line(2,-1){4.00}}
\put(79.00,10.00){\line(-2,-1){4.00}}
\put(69.00,10.00){\line(2,-1){4.00}}
\put(69.00,10.00){\line(2,1){4.00}}
\put(82.00,12.00){\line(2,-1){4.00}}
\put(86.00,10.00){\line(-2,-1){4.00}}
\put(50.00,10.00){\makebox(0,0)[cc]{\large $M^{(2)} =$}}
\end{picture}
\end{eqnarray}
where the lines  with two ingoing and outgoing arrows denote the
"anomalous" GF  $F^{(1)}$ and $F^{(2)}$ which are proportional to the gap.
Pairing phonons are not included here because their contribution is probably
small.

It is implied in the following that the initial quantities of our problem
are "observed" mean field described by a phenomenological potential,
e.g. by the Woods-Saxon one, and pairing gap. The corresponding
single-particle levels should be extracted from the observed excitation
energies of non-magic nuclei (we will perform this procedure). The initial pairing
gaps values are taken from experiment or solution of the BCS gap equation
with the phenomenologically determined pp-interaction.

 So far as the nuclear pairing problem is solved within the BCS
approach as a rule, the quasiparticle-phonon interaction is not considered
explicitely and quantitatively on this level. If we  consider explicitely the
contribution of the graphs (\ref{M1}) we must also take them into account
in the pairing problem and , therefore, to avoid double counting,
change the phenomenological pp-interaction entering the usual BCS
problem. This question will be considered elsewhere but here we use a
simplier phenomenological procedure of "refining" the gap values
from the quasiparticle-phonon interaction under consideration.
It is analogous to refining the phenomenological
single-particle energies in magic nuclei ~\cite{r6},~\cite{r7}.

In this work we study the role of the terms (\ref{M1}) and of complete taking into
account ground state correlations to describe the excitations of
$^{119}Sn$ and $^{121}Sn$. The important role of ground state correlations
was shown long ago, e.g. for the M1 excitations in $^{40}Ca$~\cite{r7} and $^{96}Zr$~\cite{r8}.
For the excitations of odd non-magic nuclei this was considered in ~\cite{r9}
but without any "refining" procedure for the case of the phenomenological
mean field. The above-mentioned procedure of "refining" the gap and
single-particle energies' values will be also developed and realized here.

\section{Equations for one-particle Green functions in non-magic nuclei}

The general system of exact equations for the
"normal" GF's G and $G^{(h)}$ and "anomalous" GF's $F^{(1)}$ and $F^{(2)}$
in a  Fermi system with pairing
has the form ~\cite{r10}:
\begin{eqnarray}
\label{green}
G=G_{0}+ G_{0}\Sigma G- G_{0}\Sigma^{(1)} F^{(2)},\\
\nonumber
F^{(2)}=G_{0}^{(h)} \Sigma^{(h)} F^{(2)} + G_{0}^{(h)} \Sigma^{(2)} G ,
\end{eqnarray}
where $G_{0}$ and $G_{0}^{(h)}$ are free GFs, i.e. the GFs of ideal gas,
$\Sigma$, $\Sigma^{(1)}$, $\Sigma^{(2)}$ and $\Sigma^{(h)}$ are
 full irreducible self-energy parts (mass operators).
Eq.(~\ref{green}) should be supplemented by the equations for
$G^{(h)}$ and $F^{(1)}$.
We use the symbolic form of equations very often.

It is very natural to single out explicitely known components of the mass
operators. So we write
\begin{eqnarray}
\label{mass}
\Sigma(\varepsilon)=\tilde \Sigma + M(\varepsilon), \hspace{0.2cm}
\Sigma^{(h)}(\varepsilon)=\tilde \Sigma^{(h)} + M^{(h)}(\varepsilon),
\\
\nonumber
\Sigma^{(1)}(\varepsilon)=\tilde \Sigma^{(1)} + M^{(1)}(\varepsilon) \equiv
\tilde \Delta^{(1)} + M^{(1)}(\varepsilon),
\\
\nonumber
\Sigma^{(2)}(\varepsilon)=\tilde \Sigma^{(2)} + M^{(2)}(\varepsilon) \equiv
\tilde \Delta^{(2)} + M^{(2)}(\varepsilon).
\end{eqnarray}
Here the first terms do not depend on the energy variable $\varepsilon$.
The quantities $\tilde \Sigma$, $\tilde \Sigma^{(h)}$ correspond to
a mean field and $\tilde \Sigma^{(1)}$, $\tilde \Sigma^{(2)}$ describe
a pairing of the BCS type. The quantities M, $M^{(1)}$, $M^{(2)}$, $M^{(h)}$
(called further as $M^{(i)}$) are not defined so far. We mean that they contain
the quasiparicle-phonon interaction. Due to the fact that, as was mentioned before,
the single-particle energies $\varepsilon_{\lambda}$ and gaps $\Delta_{\lambda}$
are phenomenological, the quantities $M^{(i)}$ should give a
contribution to $\varepsilon_{\lambda}$ and $\Delta_{\lambda}$.
Therefore, to avoid the double counting of the quasiparticle-phonon
interaction, we must "refine" the first terms of the sums (\ref{mass}) and to
obtain (see below) the corresponding $\tilde \varepsilon_{\lambda}$ and  $\tilde \Delta_{\lambda}$
from $\varepsilon_{\lambda}$ and $\Delta_{\lambda}$. These "refined"
quantities are denoted by "tilde".

Taking Eqs. (\ref{mass}) into account the general system (\ref{green})
can be transformed to the following equations (see the derivation in
~\cite{r4,r5}):
\begin{eqnarray}
\label{solv}
G = \tilde{G} +  \tilde{G}MG - \tilde{F}^{(1)}M^{(h)}F^{(2)} -  \tilde{G}M^{(1)}F^{(2)} - \tilde{F}^{(1)}M^{(2)}G
\\
\nonumber
F^{(2)} = \tilde{F}^{(2)} + \tilde{F}^{(2)}MG + \tilde{G}^{(h)}M^{(h)}F^{(2)} - \tilde{F}^{(2)}M^{(1)}F^{(2)} + \tilde{G}^{(h)}M^{(2)}G,
\end{eqnarray}
(and the same for $G^{(h)}$ and $F^{(1)}$).
The bare GF $\tilde{G}$, $\tilde{G}^{(h)}$ and $\tilde F^{(1)}$, $\tilde F^{(2)}$
are the known GFs of Gorkov. In the $\lambda$ representation of single-particle
wave functions they have the form:
\begin{eqnarray}
\label{bogol}
\tilde{G}_{\lambda}(\varepsilon) = \tilde{G}_{\lambda}^{(h)}(- \varepsilon)=
\frac{\tilde u^{2}_{\lambda}}{\varepsilon - \tilde E_{\lambda}+i\delta}+
\frac{\tilde v^{2}_{\lambda}}{\varepsilon + \tilde E_{\lambda}-i\delta}
\\
\nonumber
\tilde F^{(1)}_{\lambda} = \tilde F^{(2)}_{\lambda} =-
\frac{\tilde \Delta_{\lambda}}{2 \tilde E_{\lambda}}
(\frac{1}{\varepsilon - \tilde E_{\lambda}+i\delta}-
\frac{1}{\varepsilon + \tilde E_{\lambda}-i\delta})
\end{eqnarray}
where $\tilde u^{2}_{\lambda} = 1- \tilde v^{2}_{\lambda} =
(\tilde E_{\lambda}+\tilde \varepsilon_{\lambda}) / (2 \tilde E_{\lambda})$,
 $\tilde E_{\lambda} = \sqrt{\tilde \varepsilon_{\lambda}^{2}+\tilde \Delta_{\lambda}^{2} } $.
The "refined" quantities $\tilde \varepsilon_{\lambda}$ and gaps $\tilde \Delta_{\lambda}$
will be determined in the next Section.

The physical meaning of Eqs.(\ref{solv}) is that we have singled out
explicitely the effects of mean field and Cooper pairing in a "refined" form.
In the hamiltonian approach, the latter corresponds to the Bogolubov's
transformation but without refining from the quasiparticle-phonon interaction
in the case of taking it into account.

Let us define now the quantities $M^{(i)}$. We will take them in the simplest
$g^{2}$ approximation of Eqs. (\ref{M}, \ref{M1}). (In ~\cite{r4} it was shown
that the dimensionless quantity $g^{2}$ is a small parameter for $^{120}Sn$,
i.e. $g^{2}<1$ for the most collective low lying $2_{1}^{+}$ and $3_{1}^{-}$
phonons.) Then the Eqs.(\ref{solv}) have the following graphical form:
\begin{eqnarray}
\nonumber
\unitlength=1.00mm
\special{em:linewidth 0.4pt}
\linethickness{0.4pt}
\begin{picture}(95.00,7.00)(-10,0)
\put(-10.00,0.00){\line(1,0){8.00}}
\put(-10.00,0.50){\line(1,0){8.00}}
\put(-5.00,0.00){\line(-2,1){4.00}}
\put(-5.00,0.00){\line(-2,-1){4.00}}
\put(1.00,0.00){\makebox(0,0)[cc]{$=$}}
\put(5.00,0.00){\line(1,0){8.00}}
\put(10.00,0.00){\line(-2,1){4.00}}
\put(10.00,0.00){\line(-2,-1){4.00}}
\put(29.00,0.00){\line(1,0){20.00}}
\put(33.00,1.00){\circle{2.00}}
\put(45.00,1.00){\circle{2.00}}
\put(34.00,3.00){\circle*{1.00}}
\put(35.00,4.00){\circle*{1.0}}
\put(36.00,5.00){\circle*{1.0}}
\put(37.00,5.50){\circle*{1.00}}
\put(38.00,6.00){\circle*{1.00}}
\put(39.00,6.20){\circle*{1.00}}
\put(40.00,6.00){\circle*{1.00}}
\put(41.00,5.50){\circle*{1.00}}
\put(42.00,5.00){\circle*{1.00}}
\put(43.00,4.00){\circle*{1.00}}
\put(44.00,3.00){\circle*{1.00}}
\put(25.00,0.00){\line(1,0){5.00}}
\put(49.00,0.00){\line(1,0){5.00}}
\put(46.00,0.5){\line(1,0){8.00}}
\put(27.00,2.00){\line(2,-1){4.00}}
\put(31.00,0.00){\line(-2,-1){4.00}}
\put(37.00,2.00){\line(2,-1){4.00}}
\put(41.00,0.00){\line(-2,-1){4.00}}
\put(47.00,2.00){\line(2,-1){4.00}}
\put(51.00,0.00){\line(-2,-1){4.00}}
\put(20.00,0.00){\makebox(0,0)[cc]{$+$}}
\put(74.00,0.00){\line(1,0){20.00}}
\put(78.00,1.00){\circle{2.00}}
\put(90.00,1.00){\circle{2.00}}
\put(79.00,3.00){\circle*{1.00}}
\put(80.00,4.00){\circle*{1.0}}
\put(81.00,5.00){\circle*{1.0}}
\put(82.00,5.50){\circle*{1.00}}
\put(83.00,6.00){\circle*{1.00}}
\put(84.00,6.20){\circle*{1.00}}
\put(85.00,6.00){\circle*{1.00}}
\put(86.00,5.50){\circle*{1.00}}
\put(87.00,5.00){\circle*{1.00}}
\put(88.00,4.00){\circle*{1.00}}
\put(89.00,3.00){\circle*{1.00}}
\put(70.00,0.00){\line(1,0){5.00}}
\put(94.00,0.00){\line(1,0){5.00}}
\put(91.00,0.50){\line(1,0){8.00}}
\put(69.00,2.00){\line(2,-1){4.00}}
\put(73.00,0.00){\line(-2,-1){4.00}}
\put(74.00,0.00){\line(2,1){4.00}}
\put(74.00,0.00){\line(2,-1){4.00}}
\put(83.00,0.00){\line(2,-1){4.00}}
\put(83.00,0.00){\line(2,1){4.00}}
\put(99.0,0.00){\line(-2,1){4.00}}
\put(99.0,0.00){\line(-2,-1){4.00}}
\put(90.50,0.00){\line(2,1){4.00}}
\put(90.50,0.00){\line(2,-1){4.00}}
\put(62.00,0.00){\makebox(0,0)[cc]{$+$}}
\end{picture}
\end{eqnarray}

\begin{eqnarray}
\nonumber
\unitlength=1.00mm
\special{em:linewidth 0.4pt}
\linethickness{0.4pt}
\begin{picture}(95.00,7.00)(-10,0)
\put(29.00,0.00){\line(1,0){20.00}}
\put(33.00,1.00){\circle{2.00}}
\put(45.00,1.00){\circle{2.00}}
\put(34.00,3.00){\circle*{1.00}}
\put(35.00,4.00){\circle*{1.0}}
\put(36.00,5.00){\circle*{1.0}}
\put(37.00,5.50){\circle*{1.00}}
\put(38.00,6.00){\circle*{1.00}}
\put(39.00,6.20){\circle*{1.00}}
\put(40.00,6.00){\circle*{1.00}}
\put(41.00,5.50){\circle*{1.00}}
\put(42.00,5.00){\circle*{1.00}}
\put(43.00,4.00){\circle*{1.00}}
\put(44.00,3.00){\circle*{1.00}}
\put(25.00,0.00){\line(1,0){5.00}}
\put(49.00,0.00){\line(1,0){5.00}}
\put(46.00,0.5){\line(1,0){8.00}}
\put(27.00,2.00){\line(2,-1){4.00}}
\put(31.00,0.00){\line(-2,-1){4.00}}
\put(34.00,2.00){\line(2,-1){4.00}}
\put(38.00,0.00){\line(-2,-1){4.00}}
\put(40.00,0.00){\line(2,1){4.00}}
\put(40.00,0.00){\line(2,-1){4.00}}
\put(45.50,0.00){\line(2,1){4.00}}
\put(45.50,0.00){\line(2,-1){4.00}}
\put(50.50,2.00){\line(2,-1){4.00}}
\put(54.50,0.00){\line(-2,-1){4.00}}
\put(17.00,0.00){\makebox(0,0)[cc]{$+$}}
\put(74.00,0.00){\line(1,0){20.00}}
\put(78.00,1.00){\circle{2.00}}
\put(90.00,1.00){\circle{2.00}}
\put(79.00,3.00){\circle*{1.00}}
\put(80.00,4.00){\circle*{1.0}}
\put(81.00,5.00){\circle*{1.0}}
\put(82.00,5.50){\circle*{1.00}}
\put(83.00,6.00){\circle*{1.00}}
\put(84.00,6.20){\circle*{1.00}}
\put(85.00,6.00){\circle*{1.00}}
\put(86.00,5.50){\circle*{1.00}}
\put(87.00,5.00){\circle*{1.00}}
\put(88.00,4.00){\circle*{1.00}}
\put(89.00,3.00){\circle*{1.00}}
\put(70.00,0.00){\line(1,0){5.00}}
\put(94.00,0.00){\line(1,0){5.00}}
\put(91.00,0.50){\line(1,0){8.00}}
\put(69.00,2.00){\line(2,-1){4.00}}
\put(73.00,0.00){\line(-2,-1){4.00}}
\put(74.00,0.00){\line(2,1){4.00}}
\put(74.00,0.00){\line(2,-1){4.00}}
\put(79.50,0.00){\line(2,-1){4.00}}
\put(79.50,0.00){\line(2,1){4.00}}
\put(88.50,0.00){\line(-2,1){4.00}}
\put(88.50,0.00){\line(-2,-1){4.00}}
\put(93.00,2.00){\line(2,-1){4.00}}
\put(97.00,0.00){\line(-2,-1){4.00}}
\put(62.00,0.00){\makebox(0,0)[cc]{$+$}}
\end{picture}
\end{eqnarray}

\begin{eqnarray}
\label{grafsolv}
\unitlength=1.00mm
\special{em:linewidth 0.4pt}
\linethickness{0.4pt}
\begin{picture}(95.00,7.00)(-10,0)
\put(-12.00,0.00){\line(1,0){10.00}}
\put(-12.00,0.50){\line(1,0){10.00}}
\put(-2.00,0.00){\line(-2,1){4.00}}
\put(-2.00,0.00){\line(-2,-1){4.00}}
\put(-12.00,0.00){\line(2,1){4.00}}
\put(-12.00,0.00){\line(2,-1){4.00}}
\put(1.00,0.00){\makebox(0,0)[cc]{$=$}}
\put(4.00,0.00){\line(1,0){10.00}}
\put(14.00,0.00){\line(-2,1){4.00}}
\put(14.00,0.00){\line(-2,-1){4.00}}
\put(4.00,0.00){\line(2,1){4.00}}
\put(4.00,0.00){\line(2,-1){4.00}}
\put(29.00,0.00){\line(1,0){20.00}}
\put(33.00,1.00){\circle{2.00}}
\put(45.00,1.00){\circle{2.00}}
\put(34.00,3.00){\circle*{1.00}}
\put(35.00,4.00){\circle*{1.0}}
\put(36.00,5.00){\circle*{1.0}}
\put(37.00,5.50){\circle*{1.00}}
\put(38.00,6.00){\circle*{1.00}}
\put(39.00,6.20){\circle*{1.00}}
\put(40.00,6.00){\circle*{1.00}}
\put(41.00,5.50){\circle*{1.00}}
\put(42.00,5.00){\circle*{1.00}}
\put(43.00,4.00){\circle*{1.00}}
\put(44.00,3.00){\circle*{1.00}}
\put(25.00,0.00){\line(1,0){5.00}}
\put(49.00,0.00){\line(1,0){5.00}}
\put(46.00,0.5){\line(1,0){8.00}}
\put(29.50,2.00){\line(2,-1){4.00}}
\put(33.50,0.00){\line(-2,-1){4.00}}
\put(24.50,0.00){\line(2,1){4.00}}
\put(24.50,0.00){\line(2,-1){4.00}}
\put(37.00,2.00){\line(2,-1){4.00}}
\put(41.00,0.00){\line(-2,-1){4.00}}
\put(47.00,2.00){\line(2,-1){4.00}}
\put(51.00,0.00){\line(-2,-1){4.00}}
\put(21.00,0.00){\makebox(0,0)[cc]{$+$}}
\put(74.00,0.00){\line(1,0){20.00}}
\put(78.00,1.00){\circle{2.00}}
\put(90.00,1.00){\circle{2.00}}
\put(79.00,3.00){\circle*{1.00}}
\put(80.00,4.00){\circle*{1.0}}
\put(81.00,5.00){\circle*{1.0}}
\put(82.00,5.50){\circle*{1.00}}
\put(83.00,6.00){\circle*{1.00}}
\put(84.00,6.20){\circle*{1.00}}
\put(85.00,6.00){\circle*{1.00}}
\put(86.00,5.50){\circle*{1.00}}
\put(87.00,5.00){\circle*{1.00}}
\put(88.00,4.00){\circle*{1.00}}
\put(89.00,3.00){\circle*{1.00}}
\put(70.00,0.00){\line(1,0){5.00}}
\put(94.00,0.00){\line(1,0){5.00}}
\put(91.00,0.50){\line(1,0){8.00}}
\put(72.00,0.00){\line(2,1){4.00}}
\put(72.00,0.00){\line(2,-1){4.00}}
\put(83.00,0.00){\line(2,-1){4.00}}
\put(83.00,0.00){\line(2,1){4.00}}
\put(99.0,0.00){\line(-2,1){4.00}}
\put(99.0,0.00){\line(-2,-1){4.00}}
\put(90.50,0.00){\line(2,1){4.00}}
\put(90.50,0.00){\line(2,-1){4.00}}
\put(62.00,0.00){\makebox(0,0)[cc]{$+$}}
\end{picture}
\end{eqnarray}

\begin{eqnarray}
\nonumber
\unitlength=1.00mm
\special{em:linewidth 0.4pt}
\linethickness{0.4pt}
\begin{picture}(95.00,7.00)(-10,0)
\put(29.00,0.00){\line(1,0){20.00}}
\put(33.00,1.00){\circle{2.00}}
\put(45.00,1.00){\circle{2.00}}
\put(34.00,3.00){\circle*{1.00}}
\put(35.00,4.00){\circle*{1.0}}
\put(36.00,5.00){\circle*{1.0}}
\put(37.00,5.50){\circle*{1.00}}
\put(38.00,6.00){\circle*{1.00}}
\put(39.00,6.20){\circle*{1.00}}
\put(40.00,6.00){\circle*{1.00}}
\put(41.00,5.50){\circle*{1.00}}
\put(42.00,5.00){\circle*{1.00}}
\put(43.00,4.00){\circle*{1.00}}
\put(44.00,3.00){\circle*{1.00}}
\put(25.00,0.00){\line(1,0){5.00}}
\put(49.00,0.00){\line(1,0){5.00}}
\put(46.00,0.5){\line(1,0){8.00}}
\put(29.50,2.00){\line(2,-1){4.00}}
\put(33.50,0.00){\line(-2,-1){4.00}}
\put(24.50,0.00){\line(2,1){4.00}}
\put(24.50,0.00){\line(2,-1){4.00}}
\put(34.00,2.00){\line(2,-1){4.00}}
\put(38.00,0.00){\line(-2,-1){4.00}}
\put(40.00,0.00){\line(2,1){4.00}}
\put(40.00,0.00){\line(2,-1){4.00}}
\put(45.50,0.00){\line(2,1){4.00}}
\put(45.50,0.00){\line(2,-1){4.00}}
\put(50.50,2.00){\line(2,-1){4.00}}
\put(54.50,0.00){\line(-2,-1){4.00}}
\put(17.00,0.00){\makebox(0,0)[cc]{$+$}}
\put(74.00,0.00){\line(1,0){20.00}}
\put(78.00,1.00){\circle{2.00}}
\put(90.00,1.00){\circle{2.00}}
\put(79.00,3.00){\circle*{1.00}}
\put(80.00,4.00){\circle*{1.0}}
\put(81.00,5.00){\circle*{1.0}}
\put(82.00,5.50){\circle*{1.00}}
\put(83.00,6.00){\circle*{1.00}}
\put(84.00,6.20){\circle*{1.00}}
\put(85.00,6.00){\circle*{1.00}}
\put(86.00,5.50){\circle*{1.00}}
\put(87.00,5.00){\circle*{1.00}}
\put(88.00,4.00){\circle*{1.00}}
\put(89.00,3.00){\circle*{1.00}}
\put(70.00,0.00){\line(1,0){5.00}}
\put(94.00,0.00){\line(1,0){5.00}}
\put(91.00,0.50){\line(1,0){8.00}}
\put(72.00,0.00){\line(2,1){4.00}}
\put(72.00,0.00){\line(2,-1){4.00}}
\put(79.50,0.00){\line(2,-1){4.00}}
\put(79.50,0.00){\line(2,1){4.00}}
\put(88.50,0.00){\line(-2,1){4.00}}
\put(88.50,0.00){\line(-2,-1){4.00}}
\put(93.00,2.00){\line(2,-1){4.00}}
\put(97.00,0.00){\line(-2,-1){4.00}}
\put(62.00,0.00){\makebox(0,0)[cc]{$+$}}
\end{picture}
\end{eqnarray}
(in our linearized version we need only two equations). This corresponds
to the approximation of 1qp + 1qp$\otimes$phonon. Eqs.(\ref{grafsolv}) take
into account the ground state correlation completely because we use two
terms ("forward and backward going graphs") both in Eq.(\ref{bogol}) and in
the formulae for all the mass operator Eq.(\ref{M}, \ref{M1}) (see ~\cite{r4}),
and, of course, in the QRPA calculations of the phonon creation amplitude g.
  In the diagonal approximation for $M_{\lambda \lambda'}^{(i)}$
the secular equation, which determines the excitation energies of an odd
nucleus with pairing, is obtained from Eqs.(\ref{grafsolv}) and has the form:
\begin{eqnarray}
\label{det}
\Xi_{\lambda}(\varepsilon) \equiv
\left|
\begin{array}{cc}
1 - \tilde{G}_{\lambda}M_{\lambda} + \tilde{F}^{(1)}_{\lambda}M^{(2)}_{\lambda} &
\tilde{F}^{(1)}_{\lambda}M^{(h)}_{\lambda} +  \tilde{G}_{\lambda}M^{(1)}_{\lambda} \\
- \tilde{F}^{(2)}_{\lambda}M_{\lambda} -  \tilde{G}^{(h)}_{\lambda}M^{(2)}_{\lambda} &
1 - \tilde{G}^{(h)}_{\lambda}M^{(h)}_{\lambda} + \tilde{F}^{(2)}_{\lambda}M^{(1)}_{\lambda}
\end{array}
\right| = 0
\end{eqnarray}
where our mass operators are given by the form
$$
M_{1}(\varepsilon) = M_{1}^{(h)}(-\varepsilon) = \sum_{s,2} (g^{s}_{12})^{2}
(\frac{\tilde u_{2}^{2}}{\varepsilon - \tilde E_{2}-\omega_{s} + i \gamma}+
\frac{\tilde v_{2}^{2}}{\varepsilon + \tilde E_{2}+\omega_{s} - i \gamma}),
$$
$$
M_{1}^{(1)}(\varepsilon) = M_{1}^{(2)}(\varepsilon) = -\sum_{s,2} (g^{s}_{12})^{2}
\frac{\tilde \Delta_{2}}{ 2 \tilde E_{2}}(
\frac{1}{\varepsilon - \tilde E_{2}-\omega_{s} + i \gamma}-
\frac{1}{\varepsilon + \tilde E_{2}+\omega_{s} - i \gamma} ),
$$
here we have simplified the notations: index $1 \equiv ({\lambda_{1}}) \equiv ({ n_{1},j_{1},l_{1},m_{1} })$,
$\omega_{s}$ is the phonon energy and $g_{12}^{s}$ is the phonon creation amplitude.

The strength of the transition to the excited state $\lambda \eta$
under consideration (spectroscopic factor) is given by
\begin{eqnarray}
\label{vich}
S_{\lambda \eta}^{\pm} = \frac{(1+q_{\lambda \eta})(E_{\lambda \eta} \pm \varepsilon_{\lambda \eta})}
{\dot \Theta_{\lambda}(E_{\lambda \eta})}
\end{eqnarray}
where
\begin{eqnarray}
\label{eln}
 E_{\lambda \eta} = \sqrt{ \varepsilon_{\lambda \eta}^{2} + \Delta_{\lambda \eta}^{2} },
\end{eqnarray}

\begin{eqnarray}
\label{epsln}
 \varepsilon_{\lambda \eta} = \frac{\tilde \varepsilon_{\lambda } + M_{(even) \lambda}(E_{\lambda \eta})}
{1 + q_{\lambda \eta}},
\Delta_{\lambda \eta}= \frac {\tilde \Delta^{(1,2)}_{\lambda} + M_{\lambda}^{(1,2)}(E_{\lambda \eta}) }
{(1 + q_{\lambda \eta})},
\end{eqnarray}
\begin{eqnarray}
\nonumber
q_{\lambda \eta} = - \frac{M_{(odd) \lambda}(E_{\lambda \eta})}{E_{\lambda \eta}},
\end{eqnarray}

\begin{eqnarray}
\label{zero}
\Theta_{\lambda}(\varepsilon)=
(\varepsilon - \tilde \varepsilon_{\lambda} - M_{\lambda}(\varepsilon))
(\varepsilon + \tilde \varepsilon_{\lambda} + M^{(h)}_{\lambda}(\varepsilon))-
(\tilde \Delta_{\lambda} + M_{\lambda}^{(1)}(\varepsilon))^{2}
\end{eqnarray}
$M_{(even)}$ and $M_{(odd)}$ are the even and odd terms of the mass operator
M~\cite{r11}. The denominator in Eq. (\ref{vich}) is the energy- derivative.
Eqs.(\ref{det}-\ref{zero}) determine the characteristics of odd
nuclei with pairing in our approximation.

\section{The refinement of phenomenological single-particle energies and gaps}

The initial general system (\ref{green}) can also be transformed to another
form (see the derivation in ~\cite{r4}). Let us introduce the GF
\begin{eqnarray}
\label{other}
\bar G = G_{0} +  G_{0} (\tilde \Sigma + M) \bar G =\tilde{\bar G} +
\tilde{\bar G}  M  \bar G
\end{eqnarray}
where $\tilde{\bar G}$ determines "refined" quasiparticle energies
$\tilde \varepsilon_{\lambda}$ (we mean Landau's quasiparticles here).
Then the system (\ref{green}) can be written as one equation for G:
\begin{eqnarray}
\label{onegreen}
 G =\bar G - \bar G \Sigma^{(1)} \bar G^{(h)} \Sigma^{(2)} G.
\end{eqnarray}

In Eq.(\ref{onegreen}) we use the approximation which is diagonal in
the single-particle index $\lambda$. Let us represent the mass operator M
as a sum of its odd $M_{(odd)}$ and even $M_{(even)}$ terms ~\cite{r11} and determine the
energies $E_{\lambda}$ of the observable quasiparticle levels as
dominant solutions of Eq.(\ref{onegreen}). Then we obtain
the general formulae which connect the observable \{$\varepsilon_{\lambda}$, $\Delta_{\lambda}$\}
and refined \{$\tilde \varepsilon_{\lambda}$, $\tilde \Delta_{\lambda}$\}
quantities:
\begin{eqnarray}
\label{clear}
 \varepsilon_{\lambda } = \frac{\tilde \varepsilon_{\lambda } + M_{(even) \lambda}(E_{\lambda})}
{1 + q_{\lambda}(E_{\lambda})},
\\
\nonumber
\Delta_{\lambda} \equiv \Delta_{\lambda}^{(1,2)} = \frac{\tilde \Delta^{(1,2)}_{\lambda} + M_{\lambda}^{(1,2)}(E_{\lambda})}
{1 + q_{\lambda}(E_{\lambda})}
\end{eqnarray}
where $E_{\lambda} = \sqrt{ \varepsilon_{\lambda}^{2} + \Delta_{\lambda}^{2}}$,
$q_{\lambda } = - {M_{(odd) \lambda}(E_{\lambda })}/{E_{\lambda}}$.
Thus, if the phenomenological quantities $\varepsilon_{\lambda}$ and $\Delta_{\lambda}$ are known
we can find the bare ones,
which enter Eqs.(~\ref{solv}), (~\ref{bogol}), (~\ref{grafsolv}),
 from the solution of the nonlinear relations (\ref{clear}).

\section{Calculations of single-particle strength in $^{119}Sn$ and $^{121}Sn$. }

At first we developed and realized the procedure to extract the
phenomenological $\varepsilon_{\lambda}$ and $\Delta_{\lambda}$ from the observed
excited quasiparticle levels $E_{\lambda}$. We used the formula
$E_{\lambda}= \bar \Delta + E_{ex}^{\lambda}$, where $\bar \Delta$ is
the odd-even difference and $E_{ex}^{\lambda}$ are the excitation energies of
$^{119}Sn$ and $^{121}Sn$, and determined the quantities $\varepsilon_{\lambda}$ and $\Delta_{\lambda}$
from an iteration procedure. This procedure was organized in such a way that the
$E_{\lambda}$ values could be reproduced by the formula
 $E_{\lambda} = \sqrt{\varepsilon_{\lambda}^{2}+\Delta_{\lambda}^{2} } $.
The $\varepsilon_{\lambda}$ and $\Delta_{\lambda}$ obtained are given in
Table 1 (for detailes see ~\cite{r4}).
\begin{table}
\caption{
Refined neutrons single-particle energies $\tilde \varepsilon_{\lambda}$ and gaps
$\tilde \Delta_{\lambda}$ for $^{120}Sn$.
}
\begin{tabular}{|c|c|c|c|c|c|c|c|}
\hline
\multicolumn{1}{|c|}{$\lambda$}&
\multicolumn{1}{|c|}{$\varepsilon_{\lambda},$}&
\multicolumn{2}{|c|}{$ \tilde \varepsilon_{\lambda},МэВ $}&
\multicolumn{1}{|c|}{$\Delta_{\lambda}$}&
\multicolumn{1}{|c|}{$ \tilde \Delta_{\lambda},$}&
\multicolumn{2}{|c|}{$ \gamma_{\lambda}$ \%}\\
\cline{3-4}
\cline{7-8}
{}&{Mev}&{21 phon.}&{3 phon.}&{MeV}&{21 phon.}&{21 phon.}&{3 phon.}\\
\hline
{2d5/2}&{-2.67}&{-3.69}&{-3.41}&{1.35}&{-0.23}&{117}&{85}\\
\hline
{1g7/2}&{-1.36}&{-3.27}&{-2.54}&{1.58}&{0.76}&{52}&{33}\\
\hline
{2d3/2}&{-0.09}&{-0.17}&{-0.24}&{1.36}&{1.09}&{20}&{18}\\
\hline
{3s1/2}&{0.13}&{0.37}&{0.20}&{1.27}&{0.89}&{26}&{16}\\
\hline
{1h11/2}&{0.42}&{0.86}&{0.91}&{1.62}&{1.41}&{13}&{9}\\
\hline
\end{tabular}

\end{table}

Further, it is necessary to perform the "refining" procedure, i.e. to find
$\tilde \varepsilon_{\lambda}$ and $\tilde \Delta_{\lambda}$ from
the known $\varepsilon_{\lambda}$ and $\Delta_{\lambda}$. To do it the
non-linear Eqs.(\ref{clear}) with our $g^{2}$ choice of $M^{(i)}$ have been
solved.
The results are given in Table 1 where $\gamma_{\lambda}$ and
$\bar \gamma$ quantities were defined as follows
\begin{eqnarray}
\label{gamma}
\gamma_{\lambda}=\frac{\Delta_{\lambda} -\tilde \Delta_{\lambda}}{\Delta_{\lambda}},\hspace{1cm}
\bar \gamma= \frac{ \sum_{\lambda} \gamma_{\lambda}(2j_{\lambda}+1)}{\sum_{\lambda}(2j_{\lambda}+1)},
\end{eqnarray}
These values give the contribution of the quasiparticle-phonon pairing
mechanism caused {\em only} by the retarded pp-interaction, i.e. by the
$M^{(1)}$, $M^{(2)}$ contribution in Eq.(\ref{clear}).
In Table 1 the calculations using 3 low-lying collective
$2_{1}^{+}$, $3_{1}^{-}$, $4_{1}^{+}$ phonons and 21 phonons
are given for 5 levels near to the Fermi surface.
The phonons were calculated within the standard theory of finite
Fermi systems with phenomenological $\varepsilon_{\lambda}$
and $\Delta_{\lambda}$ which is permissible in our $g^{2}$
approximation because we omitted the  $g^{4}$ terms.
We obtained that the $\gamma_{\lambda}$ values depend rather strongly on
$\lambda$ and the mean value $\bar \gamma = 32 \% $, which was
calculated for all the 8 levels under consideration.

At last, using the $\tilde \varepsilon_{\lambda}$ and $\tilde \Delta_{\lambda}$
values as the initial data we solved Eq.(\ref{det}) and calculated the
spectroscopic factors, Eqs. (~\ref{vich}, \ref{zero}). The results and their
comparison with the experiment~\cite{r12},~\cite{r13} are given in Table 2
for all the 5 levels below and above the Fermi surface.
We obtained a reasonable agreement
with experiment
except for the $2d5/2$ and $1g7/2$ levels in $^{121}Sn$
where strength is very fragmented and small, i.e. it was
not observed very reliably.

\begin{figure}
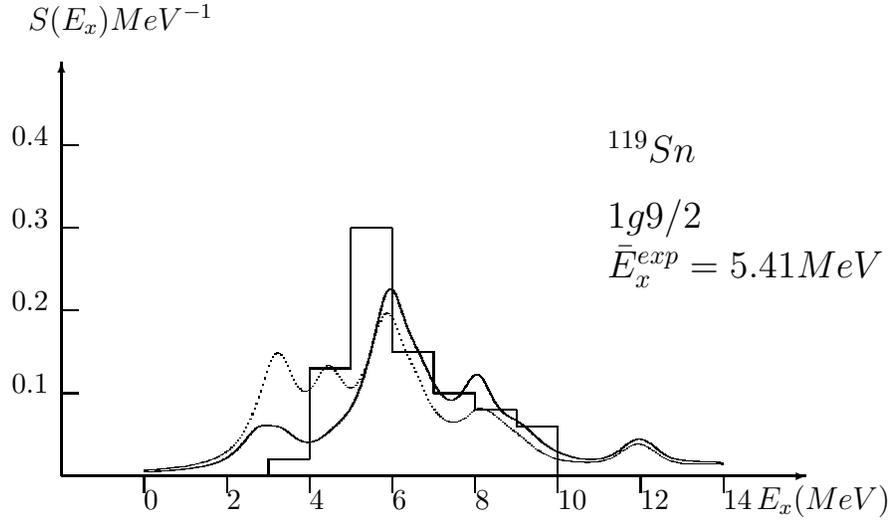

\begin{center}
\unitlength=2.2mm

\end{center}
\caption{
Strength distributions for the $1g9/2$ neutron state in $^{119}Sn$.
The solid line gives the results with taking into account all the ground state correlations,
the dashed line gives the results without them.
The averaging parameter is 1.0MeV. The experimental data (histogram) are taken
from~\cite{r12}.
}
\end{figure}

\begin{table}
\caption{
 Energies and spectroscopic factors of neutron
states in
$^{119}Sn$ (upper line) and $^{121}Sn$ (low line)$^{*}$.}
\begin{tabular}{|c|c|c|c|c|c|c|c|}
\hline
\multicolumn{1}{|c|}{$\lambda$}&
\multicolumn{3}{|c|}{$E_{x},MeV$}&
\multicolumn{3}{|c|}{$S_{\lambda}$}&
\multicolumn{1}{|c|}{$v_{\lambda}^{2}$}\\
\cline{2-4}
\cline{5-7}
{}&{Exp.}&{GSC$+$}&{GCS$-$}&{Exp.}&{GSC$+$}&{GCS$-$}&{$u_{\lambda}^{2}$}\\
\hline
{$2p^{\&}$}&{7.10}&{7.50}&{6.91}&{2.42}&{2.13}&{3.56}&{1}\\
{}&{}&{}&{}&{(4.3-10.2)}&{}&{}&{}\\
\cline{2-8}
{}&{}&{}&{}&{}&{}&{}&{0}\\
\hline
{$1g9/2$}&{5.41}&{5.90}&{5.40}&{0.53}&{0.43}&{0.43}&{$\approx 1$}\\
{}&{}&{}&{}&{(3.9-6.5)}&{}&{}&{}\\
\cline{2-8}
{}&{}&{}&{}&{}&{}&{}&{$\approx 0$}\\
\hline
{$2d5/2$}&{1.10}&{1.69}&{1.69}&{0.43}&{0.46}&{0.3}&{$ 0.95$}\\
\cline{2-8}
{}&{1.14}&{1.41}&{1.66}&{0.11}&{0.04}&{0.02}&{$0.05$}\\
{}&{}&{}&{}&{(0.0-2.0)}&{}&{}&{}\\
\hline
{$1g7/2$}&{0.79}&{1.81}&{0.95}&{0.75;0.6}&{0.66}&{0.77}&{0.83}\\
{}&{}&{}&{}&{}&{(0.0-3.0)}&{}&{}\\
\cline{2-8}
{}&{2.85}&{1.23}&{0.94}&{0.15}&{0.04}&{0.16}&{0.17}\\
{}&{}&{}&{}&{(0.0-3.5)}&{}&{}&{}\\
\hline
{$2d3/2$}&{0.02}&{0.01}&{0.02}&{0.4;0.45}&{0.4}&{0.48}&{0.52}\\
\cline{2-8}
{}&{0}&{$\approx 0$}&{$\approx 0$}&{0.44;0.65}&{0.35}&{0.43}&{0.48}\\
\hline
{$3s1/2$}&{0}&{$\approx 0$}&{$\approx 0$}&{0.26;0.32}&{0.36}&{0.42}&{0.45}\\
\cline{2-8}
{}&{0.06}&{$\approx 0$}&{$\approx 0$}&{0.3}&{0.43}&{0.51}&{0.55}\\
\hline
{$1h11/2$}&{0.09}&{0.18}&{0.15}&{0.29}&{0.26}&{0.34}&{0.37}\\
\cline{2-8}
{}&{$0.06$}&{0.17}&{$\approx0.$}&{0.49}&{0.43}&{0.56}&{0.63}\\
\hline
{$2f7/2$}&{}&{}&{}&{}&{}&{}&{$\approx 0$}\\
\cline{2-8}
{}&{2.83}&{3.11}&{2.7}&{0.35}&{0.39}&{0.59}&{$\approx 1$}\\
{}&{}&{}&{}&{(0.0-4.0)}&{}&{}&{}\\
\hline
{$3p3/2$}&{}&{}&{}&{}&{}&{}&{$\approx 0$}\\
\cline{2-8}
{}&{3.73}&{4.97}&{4.9}&{0.54}&{0.42}&{0.4}&{$1$}\\
{}&{}&{}&{}&{(0.0-5.0)}&{}&{}&{}\\
\hline
{$1h9/2$}&{}&{}&{}&{}&{}&{}&{$0$}\\
\cline{2-8}
{$+1i13/2^{(\&)}$}&{7.5.}&{7.47}&{6.61}&{$\approx 24$}&{19.92}&{22.23}&{1}\\
\hline
\end{tabular}
\\
\noindent
$^{*}${For the states without strong dominancy the energy intervals of
summation are given. The energies $E_{x}$ correspond to the
dominant peak position or to the mean energy.}
\\
$^{\&}${The quantities $S_{\lambda}*(2j+1)$ are given.}
\end{table}

In Figs.1,2 and Table 2 the results "GSC--" obtained without taking into account
all the ground state correlations (except for the QRPA ones in our
phonons) are also given. We see that the difference is rather noticeable.
The inclusion of GSC mostly gives a correct trend, both for the levels
near to the Fermi surface and for those beyond this surface, and
improves the agreement with experiment.
One can see from Fig.1,2 that it is more manifested for the
"differential" results than for the integral ones presented in Table 2.

\begin{figure}
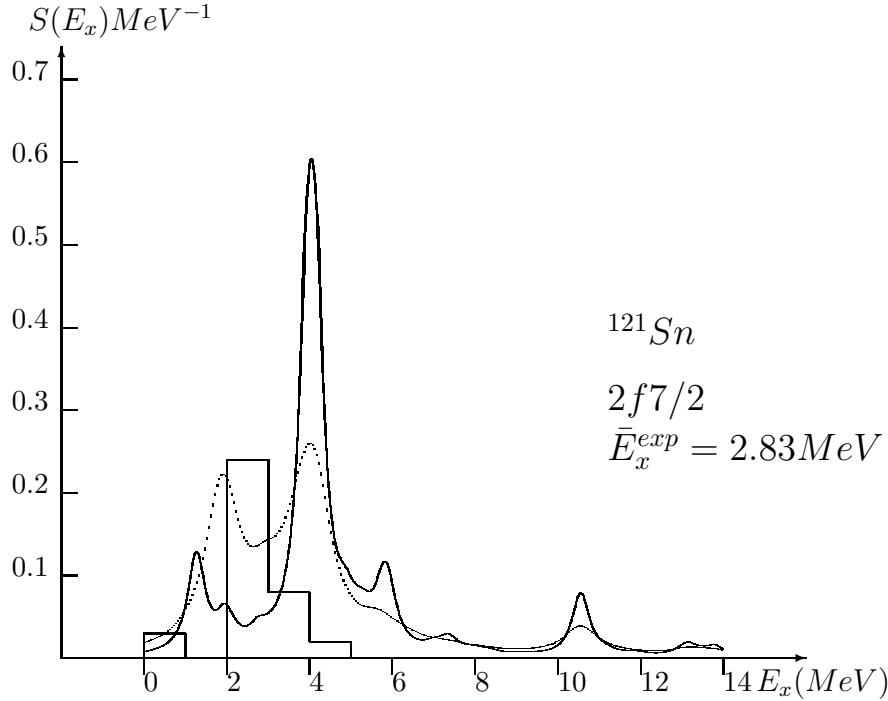

\begin{center}
\unitlength=2.2mm

\end{center}
\caption{
The same as in Fig.1 but for the $2f7/2$ neutron state in $^{121}Sn$
The experimental data (histogram) taken from~\cite{r13}.
}
\end{figure}

We also investigated the role of the terms $M^{(1)}$, $M^{(2)}$ Eq.(~\ref{M1}) of the
anomalous mass operators both for the ground state ("refining" the gap) and
for the description of excited states. It turned out that  they
give small contributions only for
states corresponding to the single-particle levels which are very
far from the Fermi surface. For the rest of the states considered,
they have a contribution and improve the agreement with the experimental
data rather often.

In conclusion, in order to calculate the distribution of  single-particle
strength in odd nuclei with pairing we formulated the method which
takes consistently into account all ground state correlations
(within the approximation used) and new
terms, i.e. Eq.(2), of the "anomalous" mass operators which are specific for
non-magic nuclei.
For the first time the method
  includes also the refinement of the phenomenological
pairing gap values from the quasiparticle-phonon
interaction under study.
A more exact than usual receipt of determing
phenomenological single- particle energies for non- magic nuclei
was realised here.
 The first calculation for $^{119}Sn$ and $^{121}Sn$
performed in the $1qp+1qp \otimes phonon$
approximation showed
a reasonable enough agreement with the available experimental data.

The mean
value of the contribution of the quasiparticle-phonon pairing mechanism
caused {\it only} by the retarded pp-interaction
was obtained for the first time, it
is about 32$\%$ for
$^{120}Sn$. A noticeable numerical role of ground state correlations,
although not so dramatic as in~\cite{r9}, was obtained.

S.K. thanks Profs. P.von Brentano, U.Kneissl and Dr. N.Pietralla for very
useful discussions.

\end{document}